01 Paper Robust V06

# Standard meta-analysis methods are not robust

S. Stanley Young, Warren B. Kindzierski

**Abstract**

P-values or risk ratios from multiple, independent studies—observational or randomized—can be computationally combined to provide an overall assessment of a research question in meta-analysis. There is a need to examine the reliability of these methods of combination. It is typical in observational studies to statistically test many questions and not correct the analysis results for multiple testing or multiple modeling, MTMM. The same problem can happen for randomized, experimental trials. There is the additional problem that some of the base studies may be using fabricated or fraudulent data. If there is no attention to MTMM or fraud in the base studies, there is no guarantee that the results to be combined are unbiased, the key requirement for the valid combining of results. We note that methods of combination are not robust; even one extreme base study value can overwhelm standard methods of combination. It is possible that multiple, extreme (MTMM or fraudulent) results can feed from the base studies to bias the combined result. A meta-analysis of observational (or even randomized studies) may not be reliable. Examples are given along with some methods to evaluate existing base studies and meta-analysis studies.

# Introduction

It is customary to examine multiple research papers and reports using a systematic review and meta-analysis in setting policy. A meta-analysis proceeds in steps: literature is gathered (now usually done electronically), titles and abstracts are examined, papers selected, and summary statistics extracted and combined mathematically into an overall p-value or a point estimate with confidence limits using Fisher's combining of p-values, FCP (e.g., **Fisher 1950**), or the method of DerSimion and Laird, DL (**DerSimion and Laird 1986**). The process goes under the name "systematic review and meta-analysis" often now part of the title of a meta-analysis paper.

As often important decisions rest on the results, there is a need to understand the reliability of the analysis process. Our idea here is to comment on the reliability of the base studies and then to focus on the mathematical combining of summary statistics coming from the base papers. Given the nature of the base studies, is combining appropriate? It is known that the summary statistics coming from the base studies are questionable if they do not consider multiple testing and multiple modeling, MTMM or possible fraud. We note that standard methods of combining these numbers—FCP or DL—are not robust to one or more extreme values. Until remedial steps are taken both in the base studies and using some robust method of combining results, we find that most meta-analysis studies are not reliable and hence of questionable value for decision making in a policy setting.

# Fisher Combining of P-values

R. A. Fisher (e.g., see **Fisher 1950, Fisher's Method**) noted that in agricultural field trials that there might be several experiments that no one of which provided sufficient evidence to inform a decision, but that together they might be sufficient for a decision. Fisher was concerned with combining *experimental* evidence where there was randomization, local control, and other sound research practices.

Fisher's method combines p-values from each experiment into one test statistic ($X^2$) using the formula

$$X^2_{2k} \sim -2 \sum_{i=1}^{k} \ln(p_i)$$

where $p_i$ is the p-value for the $i^{th}$ hypothesis test. When the p-values tend to be small, the test statistic $X^2$ will be large, which suggests that the null hypotheses are not true.

When all the null hypotheses are true and the $p_i$'s are independent, then $X^2$ has a chi-squared distribution with $2k$ degrees of freedom, where $k$ is the number of tests being combined. This result can be used to determine the p-value for $X^2$. The distribution of $X^2$ is a chi-squared distribution for the following reason: under the null hypothesis for test $i$, the p-value $p_i$ follows a uniform distribution on the interval [0,1]. The negative natural logarithm of a uniformly distributed value follows an exponential distribution. Scaling a value that follows an exponential distribution by a factor of two yields a quantity that follows a chi-squared distribution with two degrees of freedom. Finally, the sum of *k* independent chi-squared values, each with two degrees of freedom, follows a chi-squared distribution with 2*k* degrees of freedom.

Note that the Fisher test statistic is a simple sum and as such it is not robust (i.e., not strong), as a single, small p-value can force the statistic to an extreme value. **Elston 1991** notes that any p-value less than 0.37 pushes the Fisher method toward significance.

## DerSimion and Laird combining of Risk Ratios

An alternative method for combing information from multiple studies was developed by **DerSimion and Laird 1986,** DL. Their motivating studies were randomized clinical trials, again where there are expected to be sound research practices. The thinking was that there might be several small studies, which, if combined appropriately, might enlighten decision making. The small, independent studies, being randomized, provide unbiased estimates that differ one from analysis only by chance. Their method is a weighted average of the numbers coming from the base studies.

Estimated $\nu = \Sigma\ \nu_i\ w_i$

where $w_i = \sigma_i^{-2}$ for fixed effects and $(\sigma_i^2 + \tau^2)^{-1}$ for random effects. $\tau$ is a measure of study-to-study variability.

Like any average, $\nu$ is not robust to even a single extreme value and certainly not to multiple biased estimates.

Both FCP and DL methods assume the results are coming from the same underlying process and that good study technique, randomization, blocking, blinding, etc. have been employed, which gives some assurance that the numbers coming from the individual studies are unbiased. Both FCP and DL methods require unbiased numbers from the base studies. Both assume that base studies are not using fabricated or fraudulent data.

The typical observational study asks many questions and does not correct for multiple testing and multiple modeling, MTMM, [SSY: Young 2017, Young and Kindzierski 2019, Kindierski and Young 2021.] Both FCP and DL methods come with additional conditions, independence, unbiased, etc. that are unlikely to be met by observational studies and possibly many randomized clinical trials. In effect those doing meta-analysis research on observational data usually ignore these problems and take reported study results at face value. About half of the meta-analysis studies are based on observational studies. Again, neither the FCP nor DL method is robust to even a single, extreme base study.

**DerSimion and Laird 1986** state, "A problem in pooling data we have not addressed here is that of publication bias. This problem relates to studies being executed, but not reported, usually because treatment effect has not been found." In effect, studies with p-values greater than 0.05 are much less likely to be reported.

In 1975, Greenwald noted that only 6 percent of researchers were inclined to publish a negative result, whereas 60 percent were inclined to publish a positive result—a ratio of ~10 to 1. In addition, **Simonsohn, et al**. **2014**, note that replication does not necessarily support a claim if a field of research has been subject to data manipulation, or has failed to report negative results.

P-hacking is an additional problem. Using statistical software and search algorithms on computers, multiple analyses can be executed on a data set until a small p-value is found. Once a small p-value is found, a claim can be published. **Head et al. 2015** (791 Google Scholar citations as of 5/29/2021) did an extensive computer examination of the available literature and found evidence for widespread p-hacking. They opined, "... p-hacking probably does not drastically alter scientific consensuses drawn from meta-analyses." We are not so sanguine. We note that in theory, a single p-hacked study with an extremely small p-value can lead to the statistical significance of a meta-analysis. Also, evidence we show below is on the side that in any given meta-analysis there are likely many base studies that are either p-hacked or based on fraudulent data.

**Example 1**

**Young and Kindzierski 2019**, YK, examined the reliability of association of air components (ozone, NO2, PM10, CO, SO2 and PM2.5) on heart attacks given in a meta-analysis by **Mustafic et al. 2012**. The 34 base studies used by **Mustafic et al. 2012** were observational studies. YK developed p-value plots for each air component to examine the effect heterogeneity of the statistics used by **Mustafic et al. 2012**.

A p-value plot showing evidence for a real effect should have points on the plot follow approximately a line with slope < 1, where most of the p-values are small (less than 0.05) (**Young and Kindzierski 2019**, **Kindzierski et al. 2021**). The six p-value plots exhibit a bilinear, hockey stick appearance in Figure 1 (after **Young and Kindzierski 2019**). These results appear to be a mixture: some studies support an air component effect while others appear to be random (i.e., have p-value greater than 0.05).

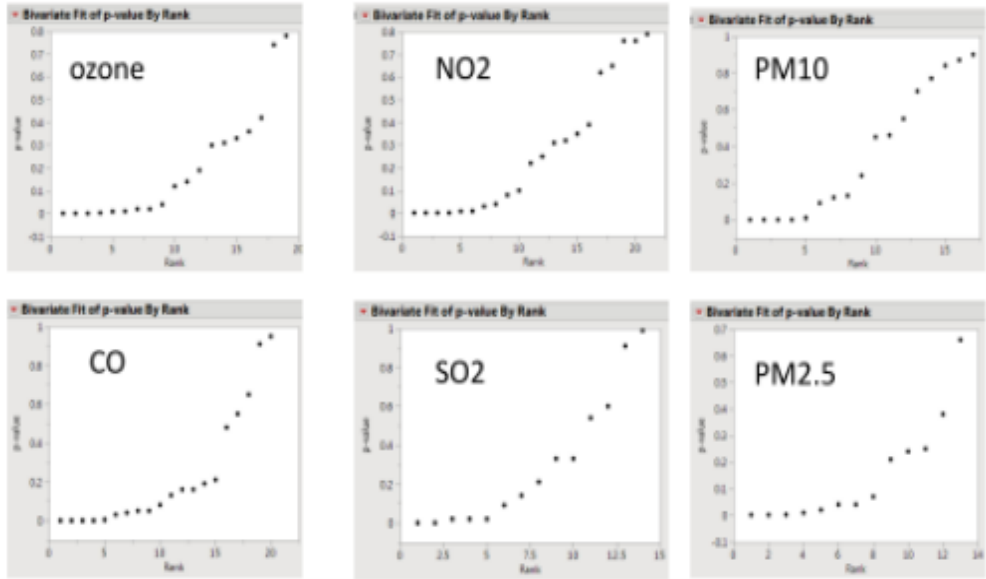

Figure 1. P-value plots for six air quality components of the case study.

YK take the position that the nominally significant p-values could well be the result of p-hacking (or some other non-treatment related reason) and the p-values on the arm of the hockey stick are valid, indicating no effect. In support of the p-hacking interpretation YK note that the analysis search spaces of the 34 base studies are large; See Table 1 (after **Young and Kindzierski 2019**).

For each base study, YK counted the number of outcomes, the number of predictors and the number of covariates. With a median of 12,288 possible analyses and an interquartile range of 2,496 to 58,368 there is considerable opportunity for p-hacking in some, if not all, of the base studies used by **Mustafic et al. 2012**. P-hacking is try this and that test/model and then selectively report results.

Table 1. Mustafic et al. 2012 authors, variable counts, and analysis search spaces for the 34 base studies, **Young and Kindzierski 2019**.

| Cit # | Author | Outcome | Predictor | Covariate | Lag | Space1 | Space2 | Space3 |
|---|---|---|---|---|---|---|---|---|
| 7 | Braga | 4 | 1 | 6 | 4 | 16 | 64 | 1,024 |
| 8 | Koken | 5 | 5 | 6 | 5 | 125 | 64 | 8,000 |
| 9 | Barnett | 7 | 5 | 10 | 1 | 35 | 1,024 | 35,840 |
| 10 | Berglind | 1 | 4 | 10 | 2 | 8 | 1,024 | 8,192 |
| 11 | Cendon | 2 | 5 | 5 | 8 | 80 | 32 | 2,560 |
| 12 | Linn | 3 | 4 | 8 | 3 | 36 | 256 | 9,216 |
| 19 | Ye | 8 | 5 | 3 | 5 | 200 | 8 | 1,600 |
| 20 | Peters | 1 | 8 | 11 | 2 | 16 | 2,048 | 32,768 |
| 21 | Rich | 1 | 5 | 9 | 6 | 30 | 512 | 15,360 |
| 22 | Sullivan | 4 | 4 | 8 | 3 | 48 | 256 | 12,288 |
| 23 | Eilstein | 1 | 12 | 5 | 6 | 72 | 32 | 2,304 |
| 24 | Lanki | 1 | 5 | 3 | 6 | 30 | 8 | 240 |
| 25 | Mate´ | 4 | 6 | 7 | 6 | 144 | 128 | 18,432 |
| 26 | Medina | 15 | 6 | 8 | 6 | 540 | 256 | 138,240 |
| 27 | Poloniecki | 7 | 5 | 5 | 1 | 35 | 32 | 1,120 |
| 28 | Stieb | 6 | 6 | 7 | 3 | 108 | 128 | 13,824 |
| 29 | Zanobetti | 5 | 2 | 11 | 3 | 30 | 2,048 | 61,440 |
| 30 | Zanobetti | 5 | 18 | 8 | 3 | 270 | 256 | 69,120 |
| 31 | Zanobetti | 5 | 2 | 9 | 2 | 20 | 512 | 10,240 |
| 32 | Hoek | 4 | 8 | 9 | 3 | 96 | 512 | 49,152 |
| 33 | Cheng | 1 | 5 | 6 | 3 | 15 | 64 | 960 |
| 34 | Hsieh | 1 | 5 | 6 | 3 | 15 | 64 | 960 |
| 35 | Pope | 1 | 2 | 13 | 7 | 14 | 8,192 | 114,688 |
| 36 | D'Ippoliti | 3 | 4 | 11 | 3 | 36 | 2048 | 73,728 |
| 37 | Henrotin | 4 | 5 | 14 | 14 | 280 | 16,384 | 4,587,520 |
| 38 | Ueda | 3 | 1 | 7 | 3 | 9 | 128 | 1,152 |
| 39 | Mann | 4 | 4 | 9 | 7 | 112 | 512 | 57,344 |
| 40 | Sharovsky | 4 | 3 | 10 | 8 | 96 | 1,024 | 98,304 |
| 41 | Belleudi | 4 | 3 | 8 | 13 | 156 | 256 | 39,936 |
| 42 | Nuvolone | 1 | 3 | 9 | 8 | 24 | 512 | 12,288 |
| 43 | Peters | 4 | 5 | 10 | 4 | 80 | 1,024 | 81,920 |
| 44 | Ruidavets | 4 | 3 | 8 | 4 | 48 | 256 | 12,288 |
| 45 | Zanobetti | 2 | 6 | 7 | 3 | 36 | 128 | 4,608 |
| 46 | Bhaskaran | 1 | 5 | 7 | 5 | 25 | 128 | 3,200 |

Note: Cit # before author name is the base study reference number; author name is first author listed; Space 1 = number of questions at issue = Outcomes × Predictors × Lags; Space 2 = $2^k$ where k = number of Covariates; Space 3 = approximation of analysis search space = Space 1 × Space 2.

## Example 2

**Vernooij et al. 2019** examined health claims associated with red and processed meat, giving multiple meta-analyses. We developed p-value plots for six health outcomes to examine the effect heterogeneity of the statistics used by **Vernooij et al. 2019** (see Figure 2). The p-value plots are for all-cause mortality, cardiovascular mortality, overall cancer mortality, breast cancer incidence, colorectal cancer incidence, and incidence of Type 2 diabetes. The six p-value plots in Figure 2a exhibit a bilinear, hockey stick appearance like Figure 1.

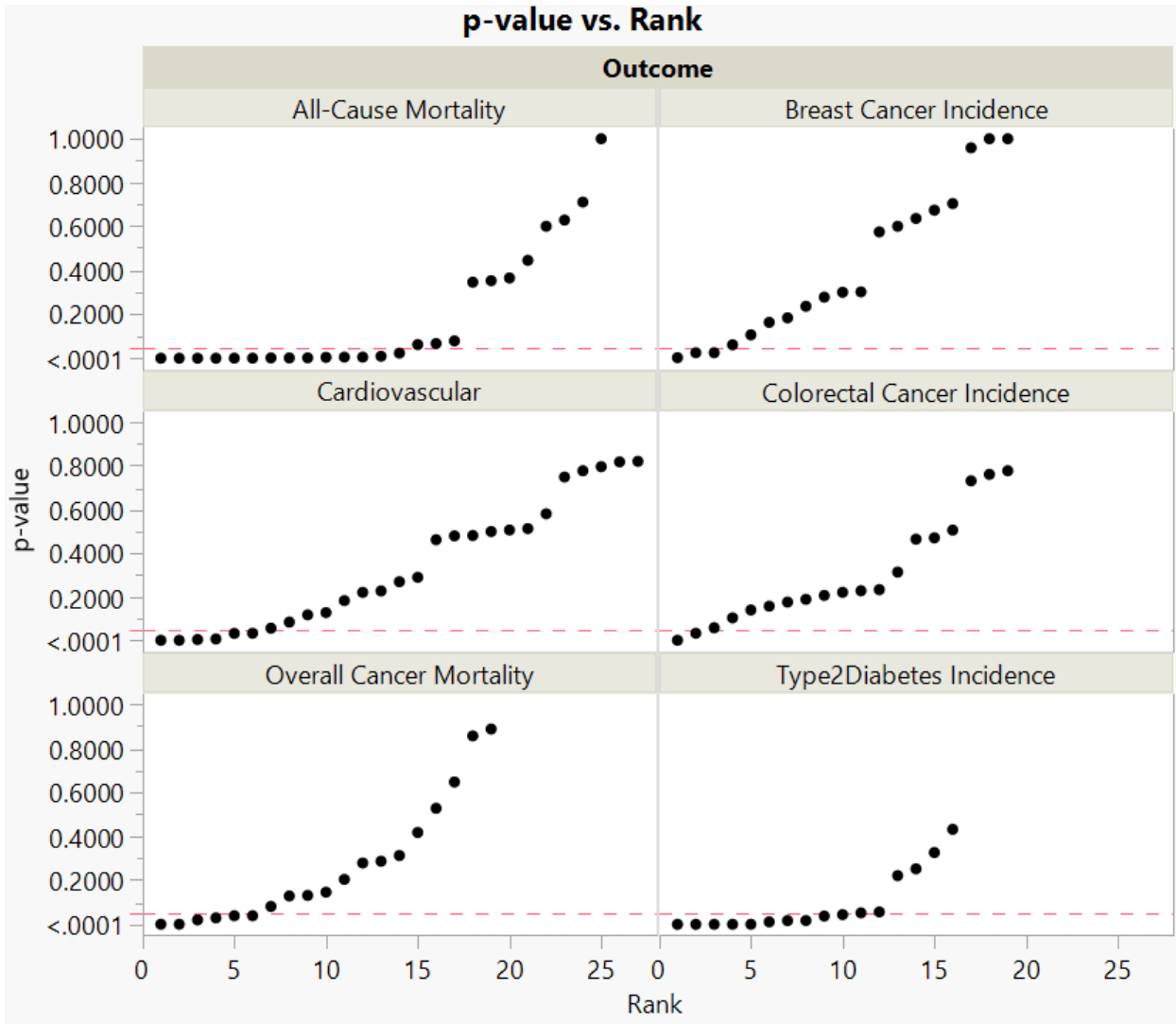

Figure 2a. P-Value plots for six health outcomes from **Vernooij et al. 2019**.

We note that the figure for Type2Diabetes is suggestive of a treatment effect. We replot that figure as Figure 2b and note that the smallest p-value is for a decrease in Type2Diabetes.

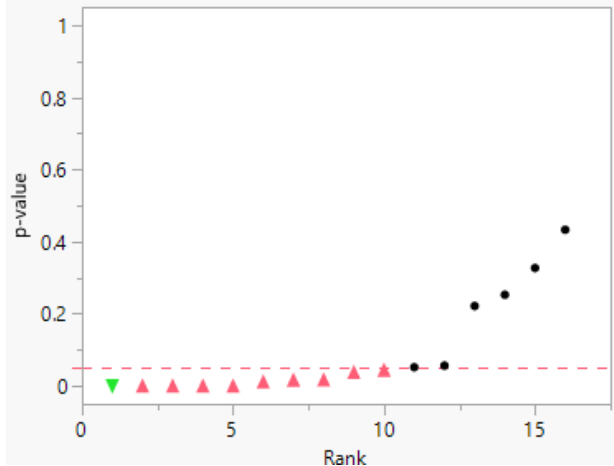

Figure 2b. Note that the smallest p-value, green down triangle, is for a decrease in Type2Diabetes.

We estimated the analysis search spaces for a random sample of 15 of the 105 used by **Vernooij et al. 2019** in their meta-analysis. The counts for outcomes, predictors, and covariates are given in Table 2. We note two things. The base studies we looked at had a food frequency questionnaire where individuals note which of many foods they ate, 12 to 280 depending on the study. We note that the base papers give results from cohort studies and also note the approximate number of papers written based on the cohort data set.

Table 2. Authors, variable counts, and analysis search spaces for a random sample of 15 of the 105 base studies used by Vernooij et al. 2019 in their meta-analysis.

| Cit # | Base Study 1st Author | Year | Foods | Outcomes | Predictors | Covariates | Space 1 | Space 2 | Space 3 |
|---|---|---|---|---|---|---|---|---|---|
| 8 | Dixon | 2004 | 51 | 3 | 51 | 17 | 153 | 131,072 | 20,054,016 |
| 31 | McNaughton | 2009 | 127 | 1 | 22 | 3 | 22 | 8 | 176 |
| 34 | Panagiotakos | 2009 | 156 | 3 | 15 | 11 | 45 | 2,048 | 92,160 |
| 38 | Héroux | 2010 | 18 | 32 | 18 | 9 | 576 | 512 | 294,912 |
| 47 | Akbaraly | 2013 | 127 | 5 | 4 | 5 | 20 | 32 | 640 |
| 48 | Chan | 2013 | 280 | 1 | 34 | 10 | 34 | 1,024 | 34,816 |
| 49 | Chen | 2013 | 39 | 4 | 12 | 5 | 48 | 32 | 1,536 |
| 53 | Maruyama | 2013 | 40 | 6 | 30 | 11 | 180 | 2,048 | 368,640 |
| 56 | George | 2014 | 122 | 3 | 20 | 13 | 60 | 8,192 | 491,520 |
| 57 | Kumagai | 2014 | 40 | 3 | 12 | 8 | 36 | 256 | 9,216 |
| 59 | Pastorino | 2016 | 45 | 1 | 10 | 6 | 10 | 64 | 640 |
| 65 | Lacoppidan | 2015 | 192 | 1 | 6 | 16 | 6 | 65,536 | 393,216 |
| 80 | Lv | 2017 | 12 | 3 | 27 | 8 | 81 | 256 | 20,736 |
| 92 | Chang-Claude | 2005 | 14 | 5 | 3 | 7 | 15 | 128 | 1,920 |
| 99 | Tonstad | 2013 | 130 | 1 | 4 | 10 | 4 | 1,024 | 4,096 |

Note: Cit# is **Vernooij et al. 2019** reference number, Author name is first author listed for reference; Year = publication year; Foods = # of foods used in Food Frequency Questionnaire; Space 1 = Outcomes × Predictors; Space 2 = $2^k$ where k = number of Covariates; Space 3 = approximation of analysis search space = Space 1 × Space 2.

We take the position that the nominally significant p-values in Figure 2 could well be the result of p-hacking (or some other non-treatment related reason including fraudulent data) and the p-values on the arm of the hockey stick are the likely valid p-values, indicating no effect. In support of the p-hacking/fraudulent data interpretation we note that the analysis search spaces of 15 randomly selected studies (Table 2) are large with a median of 20,736 possible analyses and an interquartile range of 1,536 to 368,640. We note that the existence of very small p-values, if real, would call into question non-significant p-values. Under ideal experimental conditions, **Boos and Stefanski 2011** take a p-value of 0.001 as definitive of a real effect.

## Discussion and Conclusion

The first point to make about meta-analysis studies is quality of the base studies used. Any study that has many questions under consideration and does not account for multiple testing and multiple modeling (MTMM) is not reliable. Numbers taken from such studies cannot be considered unbiased.

The first and main point of **Head et al. 2015** about p-hacking is that it is wide-spread, "while p-hacking is probably common." Our counting of outcomes, predictors, and covariates to estimate numbers of statistical tests in bases studies (e.g., refer to Tables 1 and 2) is in complete agreement. The story could end here. Until base studies are adjusted for MTMM, claims in those studies are unreliable and any meta-analysis derived from those studies are also unreliable.

Our second point is that data fabrication or fraud can come into play. The usual thinking is that there are a few "bad apples" but that mostly published papers are free from fabrication (data gardening) or fraud (making up data from whole cloth). **Redman 2013** suggest that the bad apple position deflects attention from systemic problems and gives something of a false sense of security. There are estimates of data fraud that point to rates much higher that the usual 1%, **Al-Marzouki et al. 2005, Bordewijk et al. 2020, Roberts et al. 2007 and Roberts et al. 2015**. Ten to twenty percent seems possible. Any p-values on the blade of our hockey stick figures might be from fraudulent data.

A third point is that it is obvious that if negative studies are not published then the research record is distorted and not suitable for meta-analysis. This point that has been made numerous times in literature (e.g., **DerSimion and Laird 1986**, **Egger et al. 2001**, **Sterne et al. 2001**, **Ioannidis 2008**, **NASEM 2019**). For example, **DerSimion and Laird 1986** state, "A problem in pooling data we have not addressed here is that of publication bias. This problem relates to studies being executed, but not reported, usually because a treatment effect has not been found." We agree, and again the research process could end here.

With regard to publication bias in meta-analysis, the work of **Turner et al. 2008** and **Hubbard 2015** suggest that negative studies are only being published in literature at barely more than 10% the rate of positive studies (i.e., ~1:10 for negative versus positive studies). So, a simple fold over method for analyzing publication bias in meta-analysis –which assumes 1:1 as the rate of negative versus positive studies – is not adequate for fixing the publication bias problem. Also note that positive studies are presumed to be correct by meta-analysis researchers and that is unlikely to be true (e.g., refer to **Feinstein 1988**, **Freedman 2010**, **Keown 2012**).

**Head et al. 2015** also state, "its [p-hacking] effect seems to be weak relative to the real effect sizes being measured. This statement suggests that p-hacking probably does not drastically alter scientific consensuses drawn from meta-analyses." As support **Head et al. 2015** note the presence of p-values much smaller than the marginal, yet typical cutoff value of 0.05. They presumed that the small p-values are the result of real effects. We agree with **Head et al. 2015** that small p-values will dominate in the combining of information in meta-analysis. However, we think that small p-values used in meta-analysis can be due to p-hacking or fraud and should not be presumed to be the result of real effects. Several lines of research support our interpretation. **Young and Karr 2011** noted that of 52 claims coming from *observational* studies, some with very small p-values, none replicated when examined in large, randomized trials. **Begley and Ellis 2012** noted that 47 of 53 claims coming from *experimental* biology failed to replicate. Thus, claims supported with very small p-values have often failed to replicate. These and other studies ushered in the replication crisis noted by **Baker 2016, NASEM 2019**. Note that if a small p-value is the result of p-hacking it will be accompanied with a biased treatment effect.

Current meta-analytic methods of combining (summing) information require unbiased inputs and they are not robust if this is not the case. Current literature suffers at both ends, small p-values and large p-values. Small p-values could be the result of p-hacking or fraud. Large p-values are likely missing due to publication bias. Any summation method is not robust to outliers so the combining methods will likely declare a strong signal if outliers are present. Until drastic revisions to science practice happens – i.e., p-hacking and other biases are addressed – any claims made in meta-analysis should be considered unreliable.

**References**

Al-Marzouki, S., Evans, S., Marshall, T., Roberts, I. 2005. Are these data real? Statistical methods for the detection of data fabrication in clinical trials. *BMJ (Clinical research ed.)* 331, 267. https://doi.org/10.1136/bmj.331.7511.267.

Baker, M. 2016. 1,500 scientists lift the lid on reproducibility. Nature 533, 452–454 (2016). https://doi.org/10.1038/533452a.

Begley, C. G. and Ellis, L. M. 2012. Drug development: Raise standards for preclinical cancer research. *Nature* 483, 531-33. https://doi.org/10.1038/483531a

Boos, D. D. and Stefanski L. A. 2011. P-value precision and reproducibility. *The American Statistician* 65, 213−21. https://doi.org/10.1198/tas.2011.10129.

Bordewijk, E. M., Wang, R., Aski, L. M., Gurrin, L. C., Thornton, J. G., van Wely, M., Li, W., Mol, B. W. 2020. Data integrity of 35 randomised controlled trials in women' health. The European Journal of Obstetrics & Gynecology and Reproductive Biology 249, 72–83. https://doi.org/10.1016/j.ejogrb.2020.04.016.

DerSimonian, R., and Laird, N. 1986. Meta-analysis in clinical trials. Control Clin Trials. **7**,177–188. *https://doi.org/*10.1016/0197-2456(86)90046-2.

Egger, M., Dickersin, K., Smith, G. D. 2001. Problems and Limitations in Conducting Systematic Reviews. In Systematic Reviews in Health Care: Meta-Analysis in Context, 2nd ed.; Egger, M., Smith, G.D., Altman, D.G., Eds.; BMJ Books: London, UK.

Elston, R. C. 1991. On Fisher's Method of Combining *p*-Values. Biometrical Journal 33, 339–345. https://doi.org/10.1002/bimj.4710330314.

Feinstein, A. R. 1988. Scientific standards in epidemiologic studies of the menace of daily life. Science 242, 1257−1263, https://doi.org/10.1126/science.3057627.

Fisher's Method. https://en.wikipedia.org/wiki/Fisher's_method

Fisher, R. A. 1950. Statistical Methods for Research Workers. Edinburgh: Oliver and Boyd, 11th edition, pp, 99–101.

Freedman, D.H. 2010. Lies, Damned Lies, and Medical Science; The Atlantic: Washington, DC, USA. Available online: https://www.theatlantic.com/magazine/archive/2010/11/lies-damned-lies-and-medical-science/308269/ (accessed on 10 July 2020).

Greenwald, A. G. 1975. Consequences of prejudice against the null hypothesis. *Psychological Bulletin* 82, 1: 1–20. https://doi.org/10.1037/h0076157.

Head, M. L., Holman, L, Lanfear, R., Kahn, A.T., Jennions, M. D. 2015. The extent and consequences of p-hacking in science. PLoS Biology 13, 3: e1002106. https://doi.org/10.1371/journal.pbio.1002106 .

Hubbard, R. 2015. Corrupt Research: The Case for Reconceptualizing Empirical Management and Social Science. London, UK: Sage Publications.

Ioannidis, J. P. A. 2008. Interpretation of tests of heterogeneity and bias in meta-analysis. J. Eval. Clin. Pract. 14, 951–957, https://doi.org/10.1111/j.1365-2753.2008.00986.x.

Keown, S. 2012. Biases Rife in Research, Ioannidis Says. NIH Record, Volume VXIV, No. 10. Available online: Nih-record.nih.gov/sites/recordNIH/files/pdf/2012/NIH-Record-2012-05-11.pdf (accessed on 10 July 2020).

Kindzierski. W., Young, S. S., Meyers, T., Dunn, J. 2021. Evaluation of a Meta-Analysis of Ambient Air Quality as a Risk Factor for Asthma Exacerbation. J. Respir. 1,173–196. https://doi.org/10.3390/jor1030017.

Mustafic, H., Jabre, P., Caussin, C., Murad, M. H,, Escolano, S., Tafflet, M., Perier, M-C., Marijon, E., Vernerey, D., Empana, J-P., Jouven, X. 2012. Main air pollutants and myocardial infarction: A systematic review and meta-analysis. JAMA. 307, 713−721. https://doi.org/10.1001/jama.2012.126.

NASEM (National Academies of Sciences, Engineering, and Medicine). 2019. Reproducibility and Replicability in Science; The National Academies Press: Washington, DC, USA, https://doi.org/10.17226/25303.

Redman, B. K. 2013. *Research Misconduct Policy in Biomedicine; Beyond the Bad Apple Approach*. Cambridge, MA: The MIT Press.

Roberts, I., Smith, R., Evans, S. 2007. Doubts over head injury studies. *BMJ* 334, 392. https://doi.org/10.1136/bmj.39118.480023.BE.

Roberts, I., Ker, K., Edwards, P., Beecher, D., Manno, D., Sydenham, E. 2015. The knowledge system underpinning healthcare is not fit for purpose and must change. *BMJ* 350, h2463. https://doi.org/10.1136/bmj.h2463.

Simonsohn, U., Nelson, L. D., Simmons, J. P. 2014. P-curve: A key to the filedrawer. *Journal of Experimental Psychology: General* 143, 2: 534-547. https://doi.org/10.1037/a0033242.

Sterne, J. A .C., Egger, M., Smith, G.D. 2001. Investigating and dealing with publication and other biases in meta-analysis. BMJ 323, 101–105, https://doi.org/10.1136/bmj.323.7304.101.

Turner, E. H., Matthews, A. M., Linardatos, E., Tell, R. A., Rosenthal, R. 2008. Selective publication of antidepressant trials and its influence on apparent efficacy. New England Journal of Medicine 358, 252–260. https://doi.org/10.1056/NEJMsa065779.

Vernooij, R, W. M,, Zeraatkar, D., Han, M. A., El Dib, R., Zworth, M., Milio, K., Sit, D., Lee, Y., Gomaa, H., Valli, C., Swierz, M. J., Chang, Y., Hanna, S.E., Brauer, P. M., Sievenpiper, J., de Souza, R., Alonso-Coello, P., Bala, M. M., Guyatt, G. H., Johnston, B. C. 2019. Patterns of red and processed meat consumption and risk for cardiometabolic and cancer outcomes: A systematic review and meta-analysis of cohort studies. Annals of Internal Medicine. 171, 732–742. https://doi.org/10.7326/M19-1583.

Young, S. S. 2017. Air quality environmental epidemiology studies are unreliable. Regulatory Toxicology and Pharmacology 88, 177-180. https://doi.org/10.1016/j.yrtph.2017.03.009.

Young, S.S.,and Karr, A. 2011. Deming, data and observational studies: A process out of control and needing fixing. *Significance*, September, 122-126.

Young, S.S., Kindzierski, W. B. 2019. Evaluation of a meta-analysis of air quality and heart attacks, a case study. Crit Rev Toxicol. 49, 84−95. https://doi.org/10.1080/10408444.2019.1576587.